\documentclass[preprint,aps]{revtex4}
\usepackage{mathrsfs}

\usepackage{graphicx}
\usepackage{multirow}
\usepackage{makecell}
\usepackage{booktabs}

\begin{document}

\title{Common ($\pi$,$\pi$) Band Folding and Surface Reconstruction in FeAs-Based Superconductors}

 \author{Yongqing Cai$^{1,2,\sharp}$, Tao Xie$^{1,\sharp}$, Huan Yang$^{1,2,\sharp}$, Dingsong Wu$^{1,2}$, Jianwei Huang$^{1,2}$, Wenshan Hong$^{1,2}$, Lu Cao$^{1,2}$, Chang Liu$^{1,2}$, Cong Li$^{1,2}$, Yu Xu$^{1,2}$, Qiang Gao$^{1,2}$, Taimin Miao$^{1,2}$, Guodong Liu$^{1,2,3}$, Shiliang Li$^{1,2,3}$, Li Huang$^{1}$, Huiqian Luo$^{1,2,3}$, Zuyan Xu$^{4}$, Hongjun Gao$^{1,2,3,5}$, Lin Zhao$^{1,2,3*}$ and X. J. Zhou$^{1,2,3,6,*}$}

\affiliation{
\\$^{1}$Beijing National Laboratory for Condensed Matter Physics, Institute of Physics,
Chinese Academy of Sciences, Beijing 100190, China
\\$^{2}$University of Chinese Academy of Sciences, Beijing 100049, China
\\$^{3}$Songshan Lake Materials Laboratory, Dongguan 523808, China
\\$^{4}$Technical Institute of Physics and Chemistry, Chinese Academy of Sciences, Beijing 100190, China
\\$^{5}$CAS Center for Excellence in Topological Quantum Computation, University of Chinese Academy of Sciences, Beijing 100190, China
\\$^{6}$Beijing Academy of Quantum Information Sciences, Beijing 100193, China
\\$^{\sharp}$These authors contribute equally to the present work.
\\$^{*}$Corresponding author: lzhao@iphy.ac.cn and XJZhou@iphy.ac.cn
}

\date{\today}

\begin{abstract}
High resolution angle-resolved photoemission spectroscopy (ARPES) measurements are carried out on CaKFe$_4$As$_4$, KCa$_2$Fe$_4$As$_4$F$_2$ and (Ba$_{0.6}$K$_{0.4}$)Fe$_2$As$_2$ superconductors. Clear evidence of band folding between the Brillouin zone center and corners with a ($\pi$,$\pi$) wave vector has been found from the measured Fermi surface and band structures in all the three kinds of superconductors. A dominant $\sqrt{2}\,\times\,\sqrt{2}$ surface reconstruction is observed on the cleaved surface of CaKFe$_4$As$_4$ by scanning tunneling microscopy (STM) measurements. We propose that the commonly observed $\sqrt{2}\,\times\,\sqrt{2}$ reconstruction in the FeAs-based superconductors provides a general scenario to understand the origin of the ($\pi$,$\pi$) band folding. Our observations provide new insights in understanding the electronic structure and superconductivity mechanism in iron-based superconductors.

PACS: 74. 70. -b, 74. 25. Jb, 79. 60. -i, 71. 20. -b
\end{abstract}

\maketitle


The parent compounds of the FeAs-based superconductors usually have a collinear antiferromagnetic order with a wave vector ($\pi$,$\pi$) below the magnetic transition temperature\cite{PCDai2008CCruz,XHChen2008QHuang,PCDai2009SLLi,EDagotto2012PCDai}. With doping, the antiferromagnetic order gets suppressed and antiferromagnetic fluctuation develops\cite{XHChen2009HChen,EDagotto2012PCDai,PCDai2015,JSchmalian2014RMFernandes,CMeingast2015AEBohmerNC}. The ($\pi$,$\pi$) antiferromagnetic order or fluctuation plays an important role in dictating the physical properties and superconductivity in iron-based superconductors\cite{MHDu2008IIMazin,HAoki2008KKuroki,EDagotto2012PCDai,PCDai2015,JSchmalian2014RMFernandes}. In the optimally-doped (Ba$_{0.6}$K$_{0.4}$)Fe$_2$As$_2$ superconductor, although the antiferromagnetic order is fully suppressed, the band folding connected with the wave vector ($\pi$,$\pi$) is observed\cite{SShin2017TShimojima,XJZhou2021YQCai}. The nature of the order that is responsible for the band folding remains unclear\cite{SShin2017TShimojima,XJZhou2021YQCai}. The identification of the ($\pi$,$\pi$) ordering and revelation of its origin are important in understanding the superconductivity mechanism and related physics in iron-based superconductors.

In this paper, we performed high-resolution angle-resolved photoemission spectroscopy (ARPES) on three kinds of FeAs-based superconductors, CaKFe$_4$As$_4$, KCa$_2$Fe$_4$As$_4$F$_2$ and (Ba$_{0.6}$K$_{0.4}$)Fe$_2$As$_2$. In all these samples, we observe clear signatures of band folding between the Brillouin zone center ($\Gamma$) and corners (M) connected by a ($\pi$,$\pi$) wave vector. From our scanning tunneling microscopy (STM) measurements on CaKFe$_4$As$_4$, we find that there is a $\sqrt{2}\,\times\,\sqrt{2}$ reconstruction on the As-terminated surface. Such a surface reconstruction provides a possible way to understand the ($\pi$,$\pi$) band folding in the measured FeAs-based superconductors.

The CaKFe$_4$As$_4$, KCa$_2$Fe$_4$As$_4$F$_2$ and (Ba$_{0.6}$K$_{0.4}$)Fe$_2$As$_2$ single crystals were prepared using self flux method\cite{HQLuo2018TXie,HQLuo2020WSHong,XJZhou2019JWhuang}. They all have high quality and show sharp superconducting transitions of T$_C$\,$\sim$35\,K with a width of 0.3\,K for CaKFe$_4$As$_4$\cite{HQLuo2018TXie}, T$_C$\,$\sim$33.5\,K with a width of 1\,K for KCa$_2$Fe$_4$As$_4$F$_2$\cite{XJZhou2020DSWu} and T$_C$\,$\sim$38\,K with a width of 0.5 K for (Ba$_{0.6}$K$_{0.4}$)Fe$_2$As$_2$\cite{XJZhou2019JWhuang}.
High resolution angle-resolved photoemission measurements were carried out on our lab-based ARPES system equipped with Helium discharge lamp as the light source, which can provide the photon energy of \emph{h}$\nu$ = 21.218 eV (Helium I), and  a Scienta DA30L electron energy analyzer\cite{XJZhou2008GDLiu,XJZhou2018}. The energy resolution was set at 10 meV for the Fermi surface mapping and at 4 meV for band structure measurements. The angular resolution is ~$\sim$ 0.3 $^\circ$. All the samples were cleaved \emph{in situ} at 18 K and measured in ultrahigh vacuum with a base pressure better than 5$\times$10$^{-11}$ mbar. The Fermi level is referenced by measuring on a clean polycrystalline gold that is electrically connected to the sample. High resolution STM experiments were performed at 5 K on a CreaTec ultra-high vacuum (UHV) STM/nc-AFM system with a base pressure of 2$\times$10$^{-10}$ mbar. The samples used in the STM measurements were cleaved \emph{in situ} below 77\,K and immediately transferred to an STM chamber. All STM images were obtained in constant current mode and the voltages referred to the bias on samples with respect to the tip. The tungsten tip was prepared via electrochemical etching and calibrated on a clean Au(111) substrate.

Figure 1a shows the crystal structures of CaKFe$_4$As$_4$, (Ba$_{0.6}$K$_{0.4}$)Fe$_2$As$_2$ and KCa$_2$Fe$_4$As$_4$F$_2$ superconductors. They share the same FeAs-(K,Ba,Ca)-FeAs structural unit. Among all the possible cleavage surfaces, because of the weak bonding between the FeAs layer and the adjacent (K,Ba,Ca) layer, the As-terminated surface is common for all these three kinds of superconductors. Figs. 1b-1d show the Fermi surface mapping of CaKFe$_4$As$_4$, KCa$_2$Fe$_4$As$_4$F$_2$ and (Ba$_{0.6}$K$_{0.4}$)Fe$_2$As$_2$ superconductors, respectively. These three kinds of superconductors exhibit similar Fermi surface topology, i.e., two or three hole pockets around $\Gamma$ and a small electron-like pocket around M surrounded by strong spots. These results are consistent with the previous measurements\cite{XJZhou2008LZhao,SVBorisenko2009VBZabolotnyyNature,MZHasan2008LWray,AKaminsk2016DXMou,XJZhou2020DSWu}. We note that, around the M point, in addition to the regular small electron-like pocket and surrounding strong spots, there is an extra large circle as marked by red arrows in Figs. 1b-1d. In spite of their weak spectra weight, these circles can be clearly observed, particularly in CaKFe$_4$As$_4$ (Fig. 1b) and KCa$_2$Fe$_4$As$_4$F$_2$ (Fig. 1c). We also note that the shape and size of the circles around M are similar to those of the outmost hole-like pockets around $\Gamma$.

In order to resolve the large circle feature around M more clearly and have a direct comparison with the Fermi surface around $\Gamma$, Fig. 2b shows the second derivative image of the Fermi surface mapping (Fig. 2a) of CaKFe$_4$As$_4$. We find that the extra circle around M (labelled as $\gamma'$ in Fig. 2b) is identical to the outmost hole pocket ($\gamma$) around $\Gamma$, as shown by the red solid and dashed circles with the same size. Figs. 2c-2l directly compare the band structures measured along two sets of momentum cuts (Cut 1,Cut 2) and (Cut 3,Cut 4) in both the superconducting state and normal state. All these momentum cuts are along the $\Gamma$-M$_1$ direction; Cut 1 (Cut 3) around $\Gamma$ and Cut 2 (Cut 4) around M$_2$ are connected by a ($\pi$,$\pi$) wave vector (because ($\pm\pi$,$\pm\pi$) are equivalent, we simply use ($\pi$,$\pi$) throughout the paper). To show the band structures more clearly, Figs. 2e-2h and Figs. 2k-2l show the second derivative images of the measured band structures. Along Cut 2, in addition to the usual bands around M$_2$, there appears an extra band labelled as $\gamma'$ in Fig. 2f and Fig. 2h. The position of this $\gamma'$ band relative to M$_2$ is the same as that of the $\gamma$ band relative to $\Gamma$. They also exhibit identical dispersions as shown by the red solid and dashed lines with the same shape. The same is true when comparing the band structures measured along Cut 3 (Fig. 2k) and Cut 4 (Fig. 2l). In this case, an extra band is observed along Cut 4 around M$_2$ ($\gamma'$ in Fig. 2l) which is identical to the $\gamma$ band along Cut 3 around $\Gamma$ in Fig. 2k. Figs. 2m and 2n show the comparison of MDCs at the Fermi level for the momentum cuts (Cut 1,Cut 2) and (Cut 3,Cut 4), respectively. The position of the $\gamma'$ band relative to M$_2$ observed for Cut 2 and Cut 4 is quantitatively consistent with that of the $\gamma$ band relative to $\Gamma$ observed for Cut 1 and Cut 3. The above direct comparison of the Fermi surface and band structure strongly indicates that there is a band folding between $\Gamma$ and M with a wave vector ($\pi$,$\pi$). This band folding can be observed both in the normal state and superconducting state.

We also found evidence of the ($\pi$,$\pi$) band folding in KCa$_2$Fe$_4$As$_4$F$_2$ and (Ba$_{0.6}$K$_{0.4}$)Fe$_2$As$_2$ superconductors, as shown in Fig. 3. Figs. 3a-3d directly compare the band structures measured along two momentum cuts (Cut 1 and Cut 2) in KCa$_2$Fe$_4$As$_4$F$_2$. The Cut 1 and Cut 2 are shifted relative to each other by a wave vector ($\pi$,$\pi$). Along Cut 2, there appears also an extra band labelled as $\gamma'$ in Fig. 3d. The $\gamma'$ band around M can be considered as a replica of the $\gamma$ band around $\Gamma$ because they are identical and can be connected by a ($\pi$,$\pi$) wave vector. The same is true when comparing the measured band structures in (Ba$_{0.6}$K$_{0.4}$)Fe$_2$As$_2$, as shown in Figs. 3e-3h. In this case, the $\gamma'$ band around M (Fig. 3h) is a replica of the $\gamma$ band around $\Gamma$ (Fig. 3g). The  quantitative consistency of the MDCs' peak position (marked by grey dashed lines in Figs. 2i-2j) further confirms the band folding between $\Gamma$ and M with a wave vector ($\pi$,$\pi$) in KCa$_2$Fe$_4$As$_4$F$_2$ and (Ba$_{0.6}$K$_{0.4}$)Fe$_2$As$_2$ superconductors. These observations, together with the Fermi surface folding in Fig. 1, indicate that there is a common folding effect between the Brillouin zone center and corner with a ($\pi$,$\pi$) wave vector in these FeAs-based superconductors.

In the parent and underdoped BaFe$_2$As$_2$, spin-density wave (SDW) order plays an important role in determining its electronic structure\cite{XHChen2009HChen,JSchmalian2014RMFernandes,CMeingast2015AEBohmerNC} and ARPES experiments have shown clear ($\pi$,$\pi$) band folding due to such a long-range SDW order\cite{DLFeng2009LXYang, XJZhou2009GDLiu, ZXShen2009MYi, AKaminski2010TKondo}. However, in the stoichiometric CaKFe$_4$As$_4$, KCa$_2$Fe$_4$As$_4$F$_2$ and optimally-doped (Ba$_{0.6}$K$_{0.4}$)Fe$_2$As$_2$ superconductors, the long-range SDW order is completely suppressed\cite{PCCanfield2016WRMeier,GMu2020TWang,GFChen2008JLLuo}. Therefore, the ($\pi$,$\pi$) band folding we observe in these superconductors can not be attributed to the SDW order. To the best of our knowledge, from scattering experiments, there is no report of any other long-range order in these superconductors. It is natural to ask whether the band folding can be related to the surface effect. To this end, we performed high resolution STM experiments on CaKFe$_4$As$_4$. Fig. 4a shows STM topography of CaKFe$_4$As$_4$ measured at 5\,K in a large scale (100\,nm $\times$ 100\,nm) and Fig. 4b shows a typical atomically resolved image zooming in the area marked in Fig. 4a. The bright spots form a square lattice; the distance between the two adjacent bright spots is 0.536\,nm. This distance is $\sqrt{2}$ times the lattice constant. This observation is consistent with the previous STM results\cite{HJGao2021LCao} and Fig. 4b represents As-terminated surface. In the bulk CaKFe$_4$As$_4$, the As atoms in the As layer also form a square lattice. The distance between two nearest As atoms is the same as the lattice constant (a\,=\,0.387\,nm), as shown in the left panel of Fig. 4c. On the As-terminated surface, surface reconstruction occurs, resulting in the different height and different density of states on the two adjacent As atoms\cite{HJGao2021LCao}. This renders half of the As atoms in the top As layer resolved while the other half invisible (right panel of Fig. 4c), giving rise to the $\sqrt{2}$ $\times$ $\sqrt{2}$ reconstructed structure in Fig. 4b. This $\sqrt{2}$ $\times$ $\sqrt{2}$ structure is dominant on the cleaved surface\cite{HJGao2021LCao}. This reconstruction expands the initial unit cell of a\,$\times$\,a (left panel in Fig. 4c) into $\sqrt{2}a$\,$\times$\,$\sqrt{2}a$ (right panel in Fig. 4c) in real space and, correspondingly, reduces the initial first Brillouin zone by half in the momentum space (Fig. 4d). It causes band folding between the $\Gamma$ and M points with a wave vector ($\pi$,$\pi$) (Fig. 4d), as we have observed in CaKFe$_4$As$_4$.

The $\sqrt{2}\,\times\,\sqrt{2}$ surface reconstruction is a general phenomenon observed in FeAs-based superconductors. In optimally-doped (Ba$_{0.6}$K$_{0.4}$)Fe$_2$As$_2$, the $\sqrt{2}\,\times\,\sqrt{2}$ buckling reconstructed Ba(K)-terminating layers are commonly observed on the cleaved surface\cite{SHPan2019ALi}. In KCa$_2$Fe$_4$As$_4$F$_2$, the $\sqrt{2}\,\times\,\sqrt{2}$ reconstruction of the potassium or calcium atoms is observed by STM on each cleaved surface\cite{HHWen2021WDuan}. The $\sqrt{2}\,\times\,\sqrt{2}$ surface reconstruction has also been observed in many other FeAs-based superconductors\cite{EWPlummer2009VBNascimento,MSGolden2009FMassee,VMadhavan2009FCNiestemski,JGHou2010HZhang,NKobayshi2011TNishizaki,HHWen2015DLFang}.
The common occurrence of the $\sqrt{2}\,\times\,\sqrt{2}$ surface reconstruction provides a general scenario to understand the ($\pi$,$\pi$) band folding commonly observed in the FeAs-based superconductors\cite{SShin2017TShimojima,XJZhou2021YQCai}.

In summary, we have carried out high-resolution ARPES measurements on CaKFe$_4$As$_4$, KCa$_2$Fe$_4$As$_4$F$_2$ and (Ba$_{0.6}$K$_{0.4}$)Fe$_2$As$_2$ superconductors. From the measured Fermi surface and band structures, we have found clear evidence of band folding between the $\Gamma$ and M points with a ($\pi$,$\pi$) wave vector in all the three kinds of superconductors. Our STM measurements on CaKFe$_4$As$_4$ reveal dominant $\sqrt{2}\,\times\,\sqrt{2}$ surface reconstruction on the cleaved surface. We propose that such $\sqrt{2}\,\times\,\sqrt{2}$ reconstruction, which is also commonly observed in other FeAs-based superconductors, provides a general scenario to understand the origin of the ($\pi$,$\pi$) band folding. These results provide new insights in understanding the electronic structure and superconductivity mechanism in iron-based superconductors.

\vspace{3mm}


\vspace{3mm}

\noindent {\bf Acknowledgement}\\
  We are thankful for financial support from the National Key Research and Development Program of China (Grants No. 2016YFA0300300, No. 2017YFA0302900, No. 2018YFA0704200 and No. 2019YFA0308000), the National Natural Science Foundation of China (Grants No. 11888101, No. 11922414 and No. 11874405), the Strategic Priority Research Program (B) of the Chinese Academy of Sciences (Grant No. XDB25000000), the Youth Innovation Promotion Association of CAS (Grant No. 2017013), and the Research Program of Beijing Academy of Quantum Information Sciences (Grant No. Y18G06).

\vspace{3mm}

\noindent {\bf Author Contributions}\\
 X.J.Z., L.Z. and Y.Q.C. proposed and designed the research. T.X., J.W.H., W.S.H., C.L., S.L.L. and H.Q.L. contributed in sample growth. Y.Q.C., D.S.W., J.W.H., C.L., Y.X., Q.G, T.M.M., G.D.L., Z.Y.X., L.Z. and X.J.Z. contributed to the development and maintenance of the ARPES systems and related software development. Y.Q.C. carried out the ARPES experiment with D.S.W.. H.Y., L.C., L.H. and H.J.G. carried out the STM experiment. Y.Q.C., L.Z. and X.J.Z. analyzed the data. Y.Q.C., L.Z. and X.J.Z. wrote the paper. All authors participated in discussion and comment on the paper.\\

\vspace{3mm}

\noindent{\bf Additional information}\\
Supplementary information is available in the online version of the paper.
Correspondence and requests for materials should be addressed to L. Zhao, and X. J. Zhou.

\newpage

\begin{figure*}[tbp]
\begin{center}
\includegraphics[width=0.8\columnwidth,angle=0]{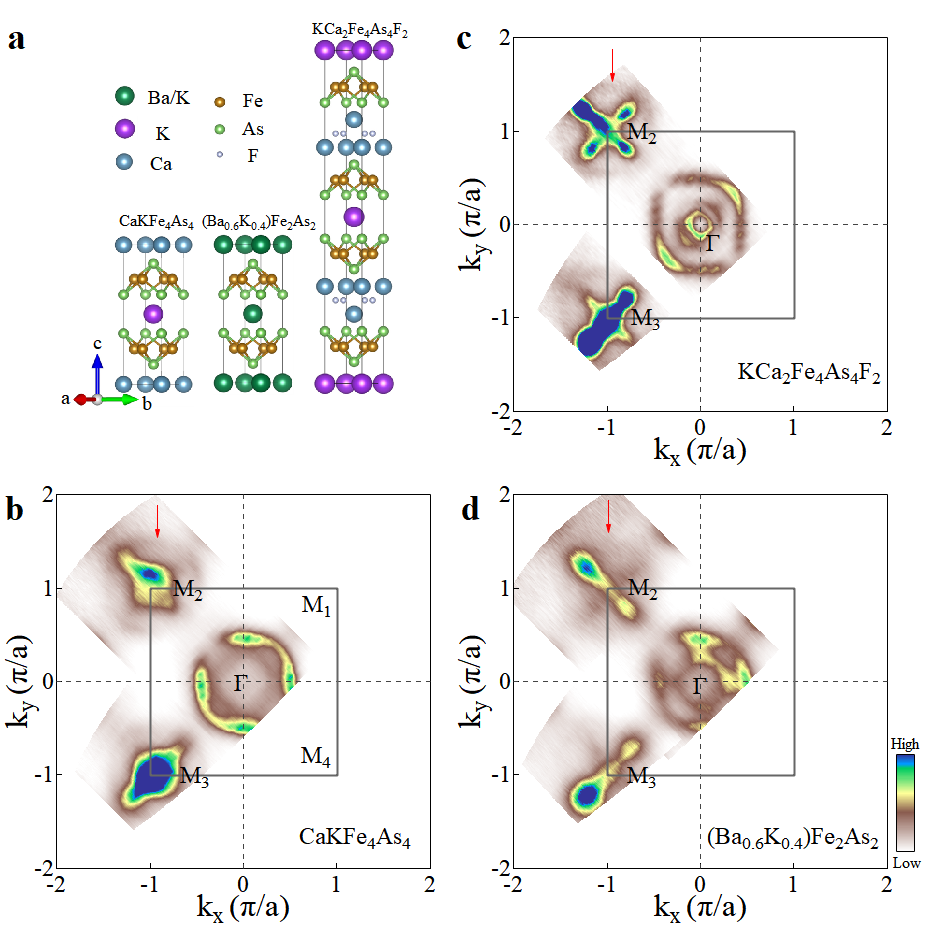}
\end{center}

\caption{{\bf Crystal structure and Fermi surface mapping of CaKFe$_4$As$_4$, KCa$_2$Fe$_4$As$_4$F$_2$ and (Ba$_{0.6}$K$_{0.4}$)Fe$_2$As$_2$ superconductors.} (a) Schematic crystal structure of CaKFe$_4$As$_4$, KCa$_2$Fe$_4$As$_4$F$_2$ and (Ba$_{0.6}$K$_{0.4}$)Fe$_2$As$_2$ Superconductors. (b-d) Fermi surface mapping of CaKFe$_4$As$_4$ measured at 18\,K (b), KCa$_2$Fe$_4$As$_4$F$_2$ measured at 38\,K (c) and (Ba$_{0.6}$K$_{0.4}$)Fe$_2$As$_2$ measured at 18\,K (d) using the Helium lamp with a photon energy of \emph{h}$\nu$=21.218\,eV. These images are obtained by integrating the spectral weight within a $\pm$ 5\,meV energy window with respect to the Fermi level E$_F$. The weak large hole-like pockets around M are marked by red arrows. For convenience, the Brillouin zone corners are labelled as $M_1$, $M_2$, $M_3$ and $M_4$.
}
\end{figure*}

\begin{figure*}[tbp]
\begin{center}
\includegraphics[width=1.0\columnwidth,angle=0]{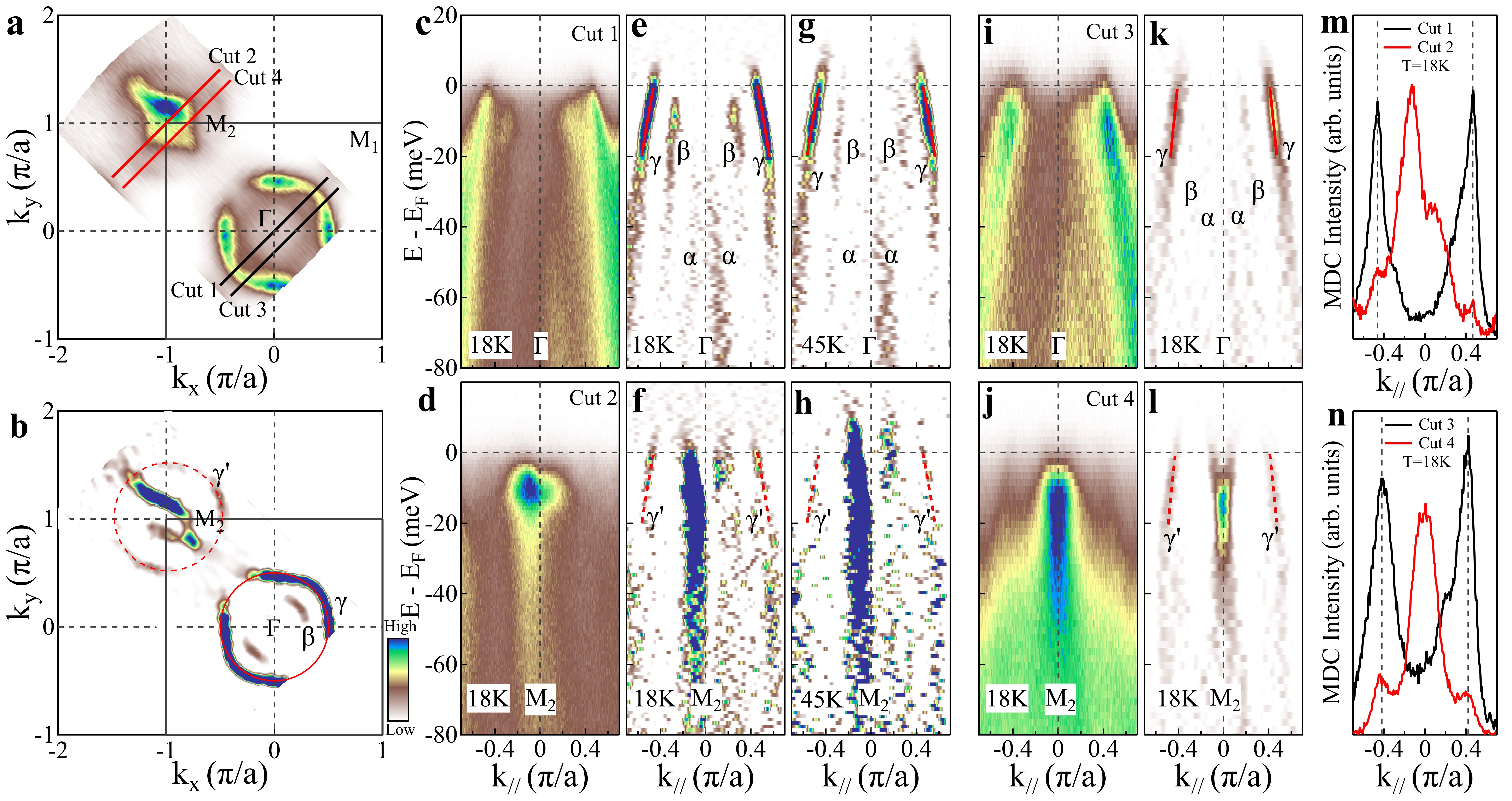}
\end{center}

\caption{{\bf Fermi surface folding and band structure folding in CaKFe$_4$As$_4$.} (a-b) Fermi surface mapping (a) and its second derivative image (b) of CaKFe$_4$As$_4$ superconductor measured at 18\,K using the Helium lamp with a photon energy of \emph{h}$\nu$=21.218 eV. These images are obtained by integrating the spectral weight within a $\pm$ 5 meV energy window with respect to the Fermi level E$_F$. The outer $\gamma$ Fermi surface sheet around $\Gamma$ is marked by a red circle. The same circle, shifted to the M point, agrees well with the additional hole-like pocket around M. (c-d) Band structure measured along Cut 1 (c) and Cut 2 (d) at 18\,K. (e-f) Corresponding second derivative images of (c-d). (g-h) Second derivative images of the band structure measured along Cut 1 (g) and Cut 2 (h) at 45\,K. (i-j) Band structure measured along Cut 3 (i) and Cut 4 (j) at 18\,K. (k-l) Corresponding second derivative images of (i-j). The location of these cuts is shown in (a). They are all along the $\Gamma$-$M_1$ direction. The observed bands around $\Gamma$ are labelled as $\alpha$, $\beta$ and $\gamma$. The outer $\gamma$ band around $\Gamma$ is marked by a red line. The same line, shifted to the M point by a wave vector ($\pi$,$\pi$), agrees well with the additional hole-like band $\gamma'$ around M. (m) MDCs at E$_F$ of the band structure measured at 18\,K for Cut 1 in (c) (black line) and Cut 2 in (d) (red line). (n) MDCs at E$_F$ of the band structure measured at 18\,K for Cut 3 in (i) (black line) and Cut 4 in (j) (red line).
}
\end{figure*}

\begin{figure*}[tbp]
\begin{center}
\includegraphics[width=1.0\columnwidth,angle=0]{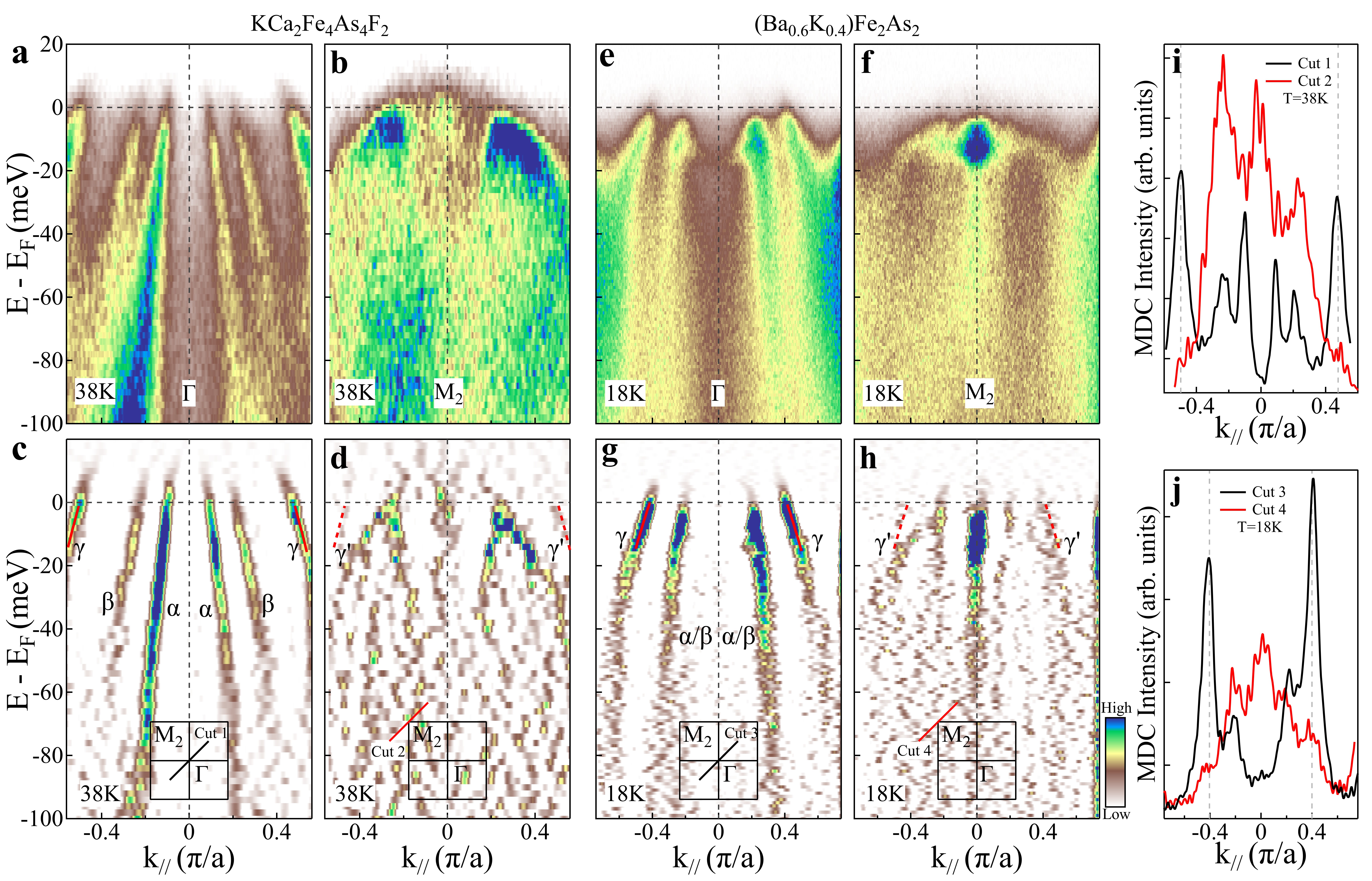}
\end{center}

\caption{{\bf Band structure folding in KCa$_2$Fe$_4$As$_4$F$_2$ and (Ba$_{0.6}$K$_{0.4}$)Fe$_2$As$_2$ superconductors.} (a-b) Band structures of KCa$_2$Fe$_4$As$_4$F$_2$ measured at 38\,K along the $\Gamma$-M$_1$ direction crossing $\Gamma$ (a) and M (b). (c-d) Corresponding second derivative images of (a-b). (e-f) Band structures of (Ba$_{0.6}$K$_{0.4}$)Fe$_2$As$_2$ measured at 18\,K along the $\Gamma$-M$_1$ direction crossing $\Gamma$ (e) and M (f). (g-h) Corresponding second derivative images of (e-f). The location of the momentum cuts is shown in the insets of (c-d) and (g-h). The inner, middle and outer bands are labelled as $\alpha$, $\beta$ and $\gamma$. The $\gamma$ band in (c) and (g) is indicated by red solid lines while $\gamma'$ (red dashed lines) in (d) and (h) represents the replica band of $\gamma$. (i) MDCs at E$_F$ of the band structure of KCa$_2$Fe$_4$As$_4$F$_2$ superconductor for Cut 1 (a) (black line) and Cut 2 (b) (red line). (j) MDCs at E$_F$ of the band structure of (Ba$_{0.6}$K$_{0.4}$)Fe$_2$As$_2$ superconductor for Cut 3 (e) (black line) and Cut 4 (f) (red line).
}
\end{figure*}

\begin{figure*}[tbp]
\begin{center}
\includegraphics[width=0.8\columnwidth,angle=0]{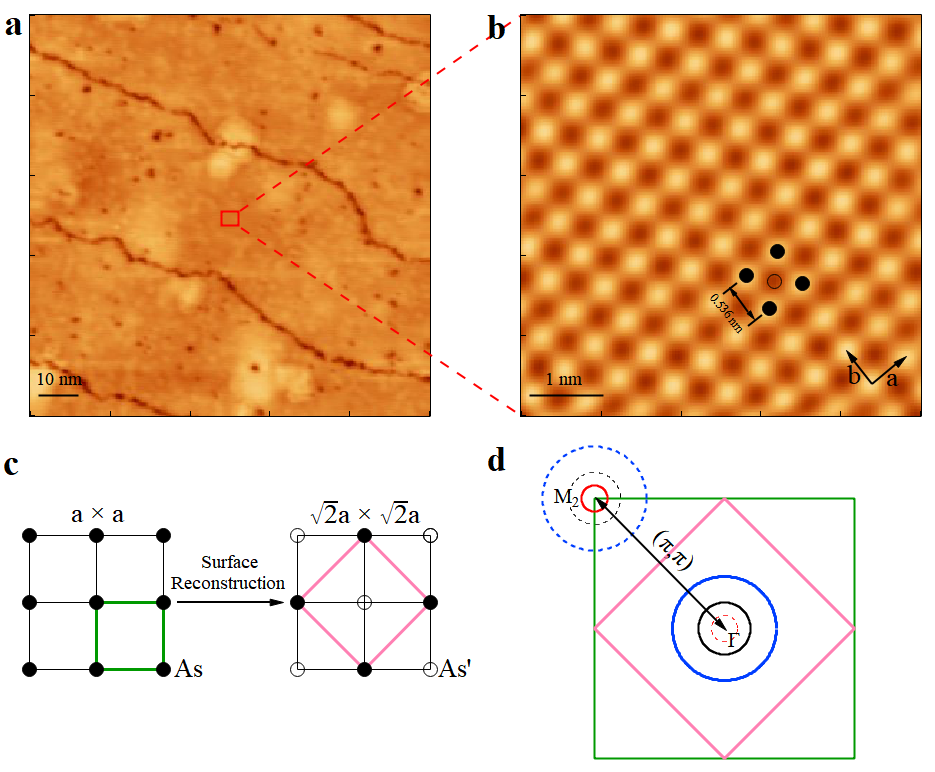}
\end{center}

\caption{{\bf STM observation of surface reconstruction in CaKFe$_4$As$_4$ and its effect on the Fermi surface folding.} (a) STM topography of the cleaved surface at 5\,K (V\,=\,–40\,mV, I\,=\,80\,pA). (b) Atomic-resolution STM image of the area marked by a red square in (a) (V\,=\,–40\,mV, I\,=\,4\,nA). The bright spots form a square lattice. The distance between two nearest bright spots is 0.536\,nm which is $\sqrt{2}$ times of the lattice constant (a\,=\,0.387\,nm). This surface is considered to be As-terminated\cite{HJGao2021LCao}. The black filled circles represent the As atoms resolved in the top layer, while black empty circles represent the position of the As atoms that are not resolved. (c) Schematic of the As layer in the bulk (left panel) and on the surface (right panel). On the surface, the two adjacent As atoms have different height with different density of states\cite{HJGao2021LCao} so that only half of the As atoms (solid circles, As) are resolved to form a $\sqrt{2}$ $\times$ $\sqrt{2}$ surface reconstruction while the other half (empty circles, As') are not observed. The green and pink squares represent the unit cell before and after the surface reconstruction, respectively. (d) Schematic of the Fermi surface folding in CaKFe$_4$As$_4$. The inner and outer hole pockets around $\Gamma$ are indicated by black and blue circles, while the electron pocket at M is indicated by the red circle. The first Brillouin zones before and after the surface reconstruction are marked by the green and pink squares, respectively. The dashed circles represents the folded Fermi surface sheets with a wave vector ($\pi$,$\pi$) between $\Gamma$ and M.
}
\end{figure*}

\end{document}